\documentclass[runningheads,citeauthoryear]{apinv}
\usepackage{epsfig,cite,graphics}
\usepackage{marvosym} %For astronomical symbols % % %
\usepackage[utf8]{inputenc}

\begin{document}

\title{Observations and light curve solutions of the\\
eclipsing stars CSS J075205.6+381909 \\
and NSVS 691550}
\titlerunning{Observations and light curve solutions of }
\author{D. Kjurkchieva\inst{1,2}, V. Popov\inst{1,2}, D. Vasileva\inst{1}, Y. Eneva\inst{1,3}, S. Ibryamov\inst{1}}
\authorrunning{Kjurkchieva, Popov, Vasileva, Eneva, Ibryamov}
\tocauthor{D. Kjurkchieva}
% Command tocautor{} is used by the Latex to give author names
% to the Contents of the volume (automatically generated)
\institute{Department of Physics and Astronomy, Shumen University,
115 Universitetska, 9700 Shumen, Bulgaria
    \and IRIDA Observatory, Rozhen NAO, Bulgaria
\and Medical University, Varna, 84 Tcar Osvoboditel str.
    \newline
    \email{d.kyurkchieva@shu.bg}    }
\papertype{Submitted on xx.xx.xxxx; Accepted on xx.xx.xxxx}
% Papertype can be "Research report", "Review", "Invited lecture", "Conference talk",
% "Conference poster", "Lecture at scientific seminar", "Summary of dissertation",  etc.
\maketitle

\begin{abstract}
The paper presents observations and light curve solutions of the
eclipsing stars CSS J075205.6+381909 and NSVS 691550. As a result
their initial epochs were determined. The target periods turned
out almost equal to the previous values. We found that NSVS 691550
is overcontact system whose components are close in temperature
while CSS J075205.6+381909 has almost contact configuration and
temperature difference of its components is around 2000 K. Both
targets undergo partial eclipses. Their stellar components seem to
obey the relations mass-temperature of MS stars.
\end{abstract}

\section*{Introduction}

Most of the W UMa stars consisting of solar-type components have
orbital periods within 0.25 d $< P <$ 0.7 d. They are recognized
by continuous brightness variations and nearly equal minima depth.
The short orbital periods of these binaries mean small orbits and
synchronized rotation and orbital revolution.

The investigation of the contact binary systems is important for
the modern astrophysics because they are natural laboratories for
study of the late stage of the stellar evolution connected with
the processes of mass and angular momentum loss, merging or fusion
of the stars (Martin {\it et al.}, 2011).

The huge surveys \emph{ROTSE, MACHO, ASAS, Super WASP, Catalina,
Kepler} increased significantly the number of stars classified as
W UMa type but small part of them are studied in details.

This paper presents our follow-up photometric observations of two
W UMa stars, CSS J075205.6+381909 (further CSS 0752+38) and NSVS
691550, and their modeling. Table~1 presents the coordinates of
our targets and available preliminary information for their light
variability from VSX database (www.aavso.org/vsx/).

\begin{table}[tp]\footnotesize
\begin{center}
\caption[]{Parameters of variability of the targets according to
VSX database
 \label{t1}}
 \begin{tabular}{cccccccc}
\hline\hline
Target        &  RA         & DEC        & mag       & ampl & $P$      & type \\
  \hline
CSS 0752+38   & 07 52 05.68 & 38 19 09.8 & 14.35(CV) & 0.17 & 0.5493   & EW \\
NSVS 691550   & 08 08 40.32 & 70 29 24.4 & 11.74(R1) & 0.26 & 0.334562 & EB/EW \\
  \hline
\end{tabular}
\end{center}
\end{table}

\section*{1. Observations}

Our CCD photometric observations of the targets in Sloan $g', i'$
bands were carried out at Rozhen Observatory with the 30-cm
Ritchey Chretien Astrograph (located into the \emph{IRIDA South}
dome) using CCD camera ATIK 4000M (2048 $\times$ 2048 pixels, 7.4
$\mu$m/pixel, field of view 40 x 40 arcmin). Information for our
observations is presented in Table~2.

\begin{table}[tp]\footnotesize
\begin{center}
\caption[]{Journal of the Rozhen photometric observations
\label{t2}}
 \begin{tabular}{cccccc}
\hline\hline
Target      &  Date        & Exposure ($g',i'$)& Number ($g',i'$) & Error ($g',i'$)\\
            &              & [sec]             &                  & [mag]     \\
  \hline
CSS 0752+38 & 2016 Feb 8   & 180, -   & 4, -   & 0.006, -  \\
            & 2016 Feb 9   & 180, 240 & 40, 40 & 0.007, 0.014\\
            & 2016 Feb 15  & 180, 240 & 63, 62 & 0.008, 0.017\\
            & 2016 Feb 28  & 180, 240 & 36, 32 & 0.009, 0.018\\
            & 2016 Mar 6   & 180, 240 & 35, 35 & 0.007, 0.016\\
            & 2016 Feb 17  & 180, 240 & 48, 48 & 0.011, 0.017\\
            & 2016 Feb 18  & 180, 240 & 50, 52 & 0.014, 0.023\\
 \hline
NSVS 691550 & 2014 Dec 19  & 60, 90   & 82, 81 & 0.003, 0.004\\
            & 2014 Dec 20  & 60, 90   &192, 187& 0.003, 0.004\\

 \hline
   \hline
\end{tabular}
\end{center}
\end{table}

\begin{table}
\begin{minipage}[t]{\textwidth}
\caption{List of the standard stars} \label{Tab3} \centering
\begin{scriptsize}
\begin{tabular}{lcccrr}
\hline\hline
Label& Star ID        & RA          & Dec          & $g'$   & $i'$   \\
  \hline
Target & CSS 0752+38  & 07 52 05.68 & +38 19 09.8  & 14.692 & 14.346 \\
Chk& UCAC4 643-044185 & 07 51 20.51 & +38 34 24.58 & 14.121 & 13.686 \\
C1 & UCAC4 643-044208 & 07 51 47.59 & +38 35 55.44 & 14.120 & 13.657 \\
C2 & UCAC4 643-044212 & 07 51 49.87 & +38 35 24.70 & 14.660 & 13.681 \\
C3 & UCAC4 643-044217 & 07 51 54.27 & +38 32 17.07 & 14.572 & 13.818 \\
C4 & UCAC4 643-044225 & 07 52 03.85 & +38 32 23.81 & 14.709 & 14.018 \\
C5 & UCAC4 643-044233 & 07 52 15.03 & +38 31 48.84 & 13.911 & 13.311 \\
C6 & UCAC4 643-044248 & 07 52 40.64 & +38 32 08.27 & 14.134 & 13.686 \\
C7 & UCAC4 643-044271 & 07 53 09.97 & +38 31 21.83 & 13.671 & 13.366 \\
C8 & UCAC4 643-044218 & 07 51 54.46 & +38 29 44.02 & 14.346 & 13.859 \\
C9 & UCAC4 643-044228 & 07 52 07.68 & +38 27 03.01 & 13.635 & 13.310 \\
C10& UCAC4 643-044235 & 07 52 19.56 & +38 25 59.02 & 14.447 & 13.755 \\
C11& UCAC4 643-044258 & 07 52 55.66 & +38 24 46.13 & 14.096 & 13.712 \\
C12& UCAC4 642-042689 & 07 52 04.30 & +38 21 25.15 & 14.479 & 13.864 \\
\hline
Target& NSVS 691550   & 08 08 40.32 & +70 29 24.4  & 11.91  & 11.280 \\
Chk& UCAC4-803-018811 & 08 08 11.10 & +70 31 24.30 & 13.893 & 12.897 \\
C1 & UCAC4-803-018819 & 08 08 35.27 & +70 27 34.33 & 12.235 & 11.135 \\
C2 & UCAC4-803-018792 & 08 07 14.92 & +70 29 38.18 & 13.617 & 13.227 \\
C3 & UCAC4-803-018788 & 08 06 53.73 & +70 24 34.73 & 13.703 & 13.357 \\
C4 & UCAC4-803-018845 & 08 09 52.06 & +70 26 47.75 & 12.788 & 12.350 \\
C5 & UCAC4-803-018826 & 08 08 56.41 & +70 25 32.22 & 13.536 & 12.918 \\
C6 & UCAC4-803-018829 & 08 09 05.15 & +70 24 34.93 & 13.981 & 13.230 \\
C7 & UCAC4-802-017856 & 08 09 29.07 & +70 23 33.50 & 13.721 & 13.218 \\
C8 & UCAC4-802-017853 & 08 09 27.67 & +70 21 39.78 & 12.754 & 12.180 \\
\hline
\end{tabular}
\end{scriptsize}
\end{minipage}
\end{table}

\begin{table}[tp]\footnotesize
\begin{center}
\caption[]{Parameters of variability of the targets according to
the Rozhen data
 \label{t4}}
 \begin{tabular}{ccccccc}
\hline\hline
Target      & $T_{0}$-2450000 & Period      & $\Delta g'$ & $\Delta i'$ & $J-K$ & T$_{m}$ \\
            &                 & [d]         & [mag]       &  [mag]      & [mag] &  [K] \\
  \hline
CSS 0752+38 & 7428.30337(7)   & 0.549284(4) & 0.283       & 0.281       & 0.306 & 6420(180) \\
NSVS 691550 & 7011.31650(4)   & 0.334562(2) & 0.338       & 0.308       & 0.419 & 5650(80) \\
  \hline
\end{tabular}
\end{center}
\end{table}

The photometric data were reduced by {\emph{AIP4WIN2.0} (Berry,
Burnell 2006). We performed aperture ensemble photometry with the
software \emph{VPHOT} using more than eight standard stars in the
observed field of each target. Table 3 presents their coordinates
and magnitudes from the catalogue UCAC4 (Zacharias {\it et al.},
2013).

We performed periodogram analysis of our data by the software
\emph{PerSea}. It led to determination of initial epochs of the
targets (Table 4) while the periods turned out almost equal to the
previous values (Table 1). The amplitudes of variability of our
observations (Table 4) are considerably larger than the
preliminary values (Table 1). This is a result of higher precision
of Rozhen observations.

\begin{figure}[!htb]
  \begin{center}
    \centering{\epsfig{file=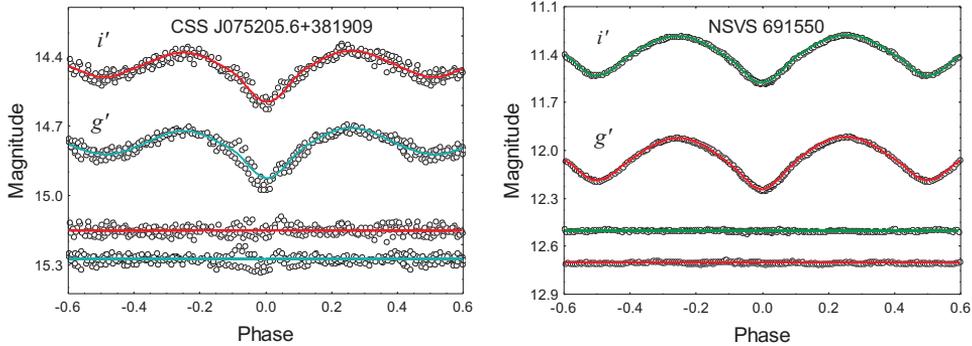, width=0.99\textwidth}}
    \caption[]{The folded light curves of targets with their fits and
the corresponding residuals (shifted vertically by different
amount to save space).}
    \label{fits}
  \end{center}
\end{figure}

\begin{figure}[!htb]
  \begin{center}
    \centering{\epsfig{file=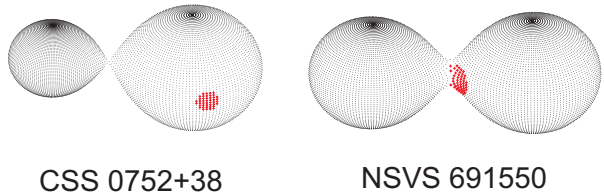, width=0.6\textwidth}}
    \caption[]{3D configurations.}
    \label{fits}
  \end{center}
\end{figure}

\section*{2. Light curve solutions}

We carried out the modeling of our data by the package
\textsc{PHOEBE} (Prsa $\&$ Zwitter 2005) based on the
Wilson--Devinney code (Wilson $\&$ Devinney 1971). It is
appropriate for our task because allows to model data in various
filters, including Sloan ones. The observational data (Fig. 1)
show that our targets are contact systems. That is why we modelled
them using the mode ''Overcontact binary not in thermal contact''.

The target temperatures $T_{m}$ were determined in advance (Table
4) on the basis of their infrared color indices \emph{(J-K)} from
the 2MASS catalog and the calibration color-temperature of
Tokunaga (2000).

The procedure of the light curve solutions consists of several
steps. Firstly, we adopted primary temperature $T_{1}$ = $T_{m}$
and assumed that the stellar components are MS stars. Then we
calculated initial (approximate) values of secondary temperature
$T_2$, mass ratio $q$, relative stellar radii $r_1$ and $r_2$,
based on the empirical relation of MS stars (Ivanov et al. 2010):
$T_2 = T_1 (d_2/d_1)^{1/4}$, $q = (T_2/T_1)^{1.7}$, $k = r_2/r_1 =
q^{0.75}$.

Further we searched for best fit varying: $T_{2}$ and $q$ around
their initial values; orbital inclination $i$ in the range
60-90$^{\circ}$ (appropriate for eclipsing stars); potentials
$\Omega_{1,2}$ in such way that the ratio $r_1/r_2$ to correspond
to the initial value $k$. We adopted coefficients of gravity
brightening 0.32 and reflection effect 0.5 appropriate for late
stars (Table 4). The limb-darkening coefficients were chosen
according to the tables of Van Hamme (1993).

After reaching the best solution (corresponding to the minimum of
$\chi^2$) we adjusted the stellar temperatures $T_{1}$ and $T_{2}$
around the value $T_m$ by the formulae (Kjurkchieva $\&$ Vasileva
2015)
\begin{equation}
T_{1}^{f}=T_{\rm {m}} + \frac{c \Delta T}{c+1}
\end{equation}
\begin{equation}
T_{2}^{f}=T_{1}^{f}-\Delta T
\end{equation}
where the quantities $c=l_2/l_1$ (the ratio of the relative
luminosities of the stellar components) and $\Delta T=T_{m}-T_{2}$
are determined from the \textsc{PHOEBE} solution.

% NSVS 691550 - O1=3.44784±0.002; O(L2)=3.022101;5370 ±13.775, ?2= 5370 ± 14

%% define your tables using this template
\begin{table}[htb]
  \begin{center}
  \caption{Fitted parameters}
  \begin{tabular}{lclll}
    \hline
Star        & $i$       &  $q$         & $T_{2}$   & $\Omega$   \\
 \hline
CSS 0752+38 & 58.5(0.2) & 0.36(0.01)   & 4477(250) & 2.59(0.02)    \\
NSVS 691550 & 60.4(0.1) & 0.845(0.001) & 5370(50)  & 3.448(0.002)    \\
\hline
  \end{tabular}
  \label{table1}
  \end{center}
\end{table}

\begin{table}[htb]
  \begin{center}
  \caption{Calculated parameters}
  \begin{tabular}{lcccccc}
    \hline
Star      & $T_{1}^{f}$ & $T_{2}^{f}$ & $r_{1}$    & $r_{2}$ & $l_{2}/l_{1}$ \\
 \hline
CSS 0752+38 & 6525(195) & 4580(250) & 0.472(0.003) & 0.294(0.004) & 0.057(0.017)   \\
NSVS 691550 & 5650(90)  & 5370(50)  & 0.404(0.002) & 0.374(0.002)
& 0.681(0.077)
  \\
\hline
  \end{tabular}
  \label{table1}
  \end{center}
\end{table}

\begin{table}[htb]
  \begin{center}
  \caption{Parameters of the surface spots}
  \begin{tabular}{ccrcl}
    \hline
star        & $\beta$     &   $\lambda$ & $\alpha$    & $k$   \\
            & [$^{\circ}$]  &  [$^{\circ}$] & [$^{\circ}$]  &       \\
 \hline
CSS 0752+38 &  90(5)      &  110(2)      &  10(1)      & 0.9(0.02)   \\
NSVS 691550 &  90(5)      &  11(2)       &  12(1)      & 0.87(0.02)   \\
\hline
  \end{tabular}
  \label{table1}
  \end{center}
\end{table}

Although \textsc{PHOEBE} works with potentials, it gives a
possibility to calculate directly all values (polar, point, side,
and back) of the relative radius $r_i=R_i/a$ of each component
($R_i$ is linear radius and \emph{a} is orbital separation).
Moreover, \textsc{PHOEBE} yields as output parameters bolometric
magnitudes $M_{bol}^i$ of the two components in conditional units
(when radial velocity data are not available). But their
difference $M_{bol}^2-M_{bol}^1$ determines the true luminosity
ratio $c=L_2/L_1=l_2/l_1$.

The formal \textsc{PHOEBE} errors of the fitted parameters were
unreasonably small. That is why we estimated the parameter errors
manually based on the following rule (Dimitrov $\&$ Kjurkchieva
2017). The error of parameter $b$ corresponded to that deviation
$\Delta b$ from its final value $b^{f}$ for which the mean
residuals increase by 3$\bar{\sigma}$ ($\bar{\sigma}$ is the mean
photometric error of the target).

Table~5 contains the final values of the fitted stellar parameters
and their uncertainties: inclination \emph{i}; mass ratio
\emph{q}; potentials $\Omega_{1, 2}$; secondary temperature
$T_{2}$. Table 6 exhibits the calculated parameters: stellar
temperatures $T_{1, 2}^f$; relative stellar radii $r_{1, 2}$ (back
values); ratio of relative stellar luminosities $l_2/l_1$. Their
errors are determined from the uncertainties of fitted parameters
used for their calculation.

The synthetic curves corresponding to the parameters of our light
curve solutions are shown in Fig. 1 as continuous lines while
Figure 2 exhibits 3D configurations of the targets. Table 7 shows
the parameters (latitude $\beta$, longitude $\lambda$, angular
size $\alpha$ and temperature factor $\kappa = T_{sp}/T_{st}$) of
the spots which were necessary to reproduce the light curve
asymmetries.

\section*{4. Analysis of the results}

The analysis of the light curve solutions led us to several
conclusions.

(a) NSVS 691550 is overcontact system while CSS 0752+38 is almost
contact binary (Fig. 2).

(b) The components of CSS 0752+38 and NSVS 691550 are F -- K
stars.

(c) The difference between the component temperatures of the
overcontact system is around 300 K, while that of CSS 0752+38
reaches almost 2000 K.

(d) The two targets undergo partial eclipses.

(e) The mass ratio of NSVS 691550 is near 0.84 while that of CSS
0752+38 is 0.35. This result means that the stellar components of
our targets almost obey the relations mass-temperature of MS
stars.

(f) Light curve asymmetries of the targets were reproduced by cool
spots on their primary components.

%CSS 0752+38(Drake et al. 2014)

%NSVS 691550 (Gettel et al. 2006)

\textbf{Acknowledgements.} This work was supported partly by
projects DN 08/20 and DM 08/02 of the Foundation for Scientific
Research of the Bulgarian Ministry of Education and Science as
well as by project RD 08-102 of Shumen University.

%\newpage


\begin{thebibliography}{}

\bibitem[2006]{ber}
Berry R., Burnell J., 2006, The Handbook of Astronomical Image
Processing with AIP4WIN2 software, Willmannn-Bell.Inc., WEB

\bibitem[2015]{dim}
Dimitrov D., Kjurkchieva D., 2015, Mon. Not. R. Astron. Soc., 448,
2890

\bibitem[2017]{dim}
Dimitrov D., Kjurkchieva D., 2017, Mon. Not. R. Astron. Soc.,
accepted

\bibitem[2010]{iv}
Ivanov V., Kjurkchieva D., Srinivasa Rao, 2010, BASI,

\bibitem[2015]{kju}
Kjurkchieva, D., Vasileva, D., 2015, PASA, 32, 23

\bibitem[2011]{mar}
Martin E.L., Spruit H.C., Tata R., 2011, A $\&$ A, 535, A50

\bibitem[2005]{prsa}
Prsa A., Zwitter T., 2005, Astrophys. J. Suppl. Ser., 628, 426

\bibitem[1992]{ruc}
Rucinski, S.M., 1992, AJ., 103, 960

\bibitem[2000]{tok}
Tokunaga A.T., 2000, Allen's astrophysical quantities, Springer,
New York

\bibitem[1993]{ham}
Van Hamme W. 1993, Astron. J., 106, 2096

\bibitem[1971]{wd}
Wilson R. E., Devinney E. J., 1971, Astrophys. J., 166, 605

\bibitem[2013]{zah}
Zacharias N. et al., 2013, AJ, 145, 44

\end{thebibliography}
\end{document}